\documentstyle[11pt,newpasp,twoside,epsf]{article}
\newcommand{\yygem}{YY\,Gem~}
\newcommand{\myrule}{\rule[-0.1cm]{0.cm}{0.5cm}}

\def\edcomment#1{\iffalse\marginpar{\raggedright\sl#1\/}\else\relax\fi}
\reversemarginpar

\begin{document}
\title{The joint {\em XMM-Newton} and {\em Chandra} view of YY\,Gem}
\author{B. Stelzer, V. Burwitz, R. Neuh\"auser}
\affil{Max-Planck Institut f\"ur extraterrestrische Physik, Postfach 1312,
 D-85741 Garching, Germany}
\author{M. Audard}
\affil{Paul Scherrer Institut, W\"urenlingen \& Villigen, 5232 Villigen PSI,
Switzerland}
\author{J. H. M. M. Schmitt}
\affil{Hamburger Sternwarte, Gojenbergsweg 112, D-21029 Hamburg, Germany}

\begin{abstract}
We have observed the flare star 
YY\,Gem simultaneously with {\em XMM-Newton} and 
{\em Chandra} as part of a multi-wavelength campaign aiming at a study
of variability related to magnetic activity in this short-period eclipsing
binary. Here we report on the first results from the analysis of the 
X-ray spectrum. The vicinity of the star provides high enough S/N
in the CCD cameras onboard {\em XMM-Newton} to allow for 
time-resolved spectroscopy. Since the data are acquired simultaneously 
they allow for a cross-calibration check of the performance of the 
{\em XMM-Newton} RGS and the LETGS on {\em Chandra}.
\end{abstract}

\section{Introduction}

YY\,Gem is the optically faintest of the three visual binaries in the
Castor sextuplet. 
It is itself an eclipsing spectroscopic binary with period of
$0.81$\,d. The two components of YY\,Gem are both of spectral type dM1e,
and belong to the class of BY\,Dra variables. Indeed, \yygem was the first
stellar system on which periodic photometric variability was detected
(Kron 1952). Since the discovery of X-ray emission from the Castor system
by the {\em Einstein} satellite, the system was studied by virtually all 
X-ray observatories (Vaiana et al. 1981, Pallavicini et al. 1990, Gotthelf
et al. 1994, Schmitt et al. 1994, G\"udel et al. 2001). 
Flares on YY\,Gem have been recorded from all parts of the electromagnetic
spectrum. The extraordinary activity of this object may be related to 
its binarity (the frequency of photometric flares seems to be enhanced in 
the interbinary space suggesting interaction between the magnetospheres of 
the two stellar components; Doyle \& Mathioudakis 1990), 
and makes it a prime target for simultaneous monitoring 
at different wavelengths.

\section{Observations and Data Analysis}

\yygem was observed by both {\em Chandra} and {\em XMM-Newton} on
Sep 29/30, 2000 for a total observing time of 59\,ksec and 55\,ksec,
respectively. The {\em XMM-Newton} observations were obtained in 
the full-frame mode of EPIC-pn, with the thick filter inserted for both
pn and MOS. We perform the data analysis 
with the standard {\em XMM-Newton} Science Analysis System (SAS). 
{\em Chandra} was used in the LETGS configuration, i.e. the Low Energy
Transmission Grating (LETG) combined with the High Resolution Camera for
Spectroscopy (HRC-S). We extracted the {\em Chandra} lightcurves and
spectra using programs written in IDL
version 5.4. The extraction areas for source and background spectrum
are those defined in the {\em Chandra} User's Guide.
  
The time of observation for the individual X-ray
instruments is given in Table~1. We display the
corresponding X-ray lightcurves in Fig.~1. The orbital phase has been
computed from the ephemeris of Torres \& Ribas (2001). First inspection 
reveals strong variability throughout the whole observation, including
two large flares, and two `high states' (i.e. extended phases of enhanced
emission) near the end of the observation. The secondary eclipse is
clearly identified as a dip in the lightcurve close to orbital phase $0.5$.
Note, that the minimum of the X-ray lightcurve is not exactly centered
on $\Phi = 0.5$, but slightly offset towards earlier times. 
As we have observed simultaneously with two independent
satellites a timing error is very unlikely. This shift may indicate 
an inhomogeneous distribution of emitting material in the coronae of the
\yygem binary.
\begin{table}
\begin{center}
\caption{Observing log for the {\em XMM-Newton} and {\em Chandra} observations
of the Castor system on Sep 29/30, 2000.}
\begin{tabular}{lrrrrr}\hline
Instrument & \multicolumn{2}{c}{UT} & \multicolumn{2}{c}{JD - 2451817} &
           Expo \\ 
           & Start & Stop & Start & Stop & [ksec] \\ \hline
\multicolumn{6}{c}{{\em XMM-Newton}}\\ \hline
EPIC-pn    & 18:56 & 08:57 & 0.2888 & 0.8730 & 50.47 \\ 
EPIC-MOS   & 18:15 & 08:51 & 0.2604 & 0.8689 & 52.57 \\
RGS        & 18:07 & 09:27 & 0.2542 & 0.8937 & 55.26 \\ \hline
\multicolumn{6}{c}{{\em Chandra}}\\ \hline
LETGS     & 21:30  & 13:54 & 0.3958 & 1.0792 & 59.00 \\ \hline
\end{tabular}
\end{center}
\end{table}
\begin{figure}
\plotfiddle{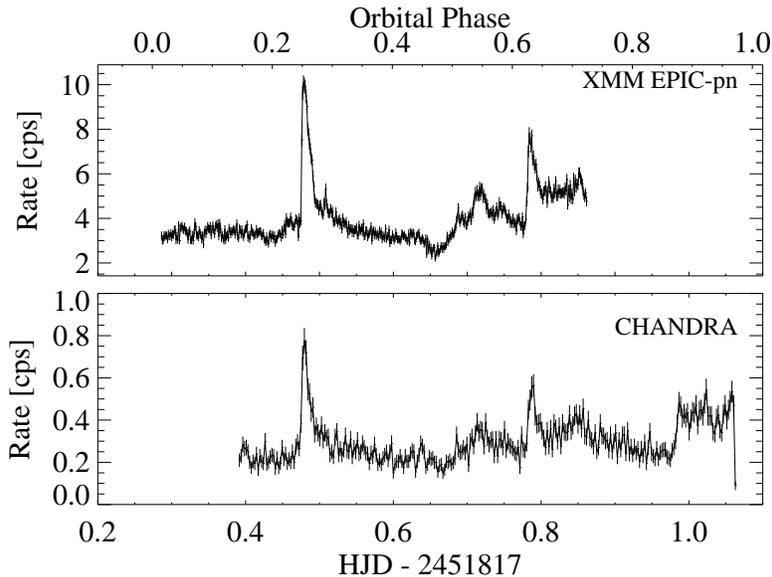}{7cm}{0}{65}{65}{-150}{0}
\caption{X-ray lightcurve of YY\,Gem observed by {\em XMM-Newton} 
(top panel) and {\em Chandra} (bottom panel) on Sep 29/30,
2000. The {\em Chandra} lightcurve has been extracted from the zeroth order
image. For {\em XMM-Newton} we show the lightcurve observed by EPIC-pn. 
The orbital ephemeris is from Torres \& Ribas (2001).}
\end{figure}

The combination of {\em XMM-Newton} and {\em Chandra} allows to examine the 
X-ray spectrum of YY\,Gem with intermediate (EPIC) 
and high (LETGS, RGS) resolution, 
and to compare the performance of the grating instruments on both satellites. 
The CCD spectra obtained with the EPIC
are analysed in the XSPEC environment (version 11.0.1). 

\section{The {\em XMM-Newton} EPIC spectrum}

We start with the analysis of the quiescent spectrum observed prior to 
the first large flare in Fig.~1 (JD 2451817.256 $-$ JD 2451817.450).  
Following G\"udel et al. (2001) we represent the quiescent EPIC spectrum
of YY\,Gem by a 3-temperature (3-T) model for thermal emission from 
an optically thin plasma (VMEKAL). In order to better constrain the 
spectral model we  
analyse the spectra from the pn and the two MOS detectors simultaneously. 
For the joint modeling of the spectrum from these three instruments 
we add a constant normalization factor to make up for uncertainties in the 
absolute calibration of the detectors. The EPIC spectrum for the pre-flare 
phase is shown in Fig.~2, and the best fit parameters from the 3-T model 
are summarized in Table~2. 
\begin{table}
\begin{center}
\caption{Spectral parameters for the quiescent state of \yygem (derived
from $t <$ JD 2451817.450). Normalization constants for cross-calibration
of the three instruments (pn, MOS\,1, MOS\,2) are: 
$N_{\rm pn} \equiv 1$ (fixed), 
$N_{\rm mos1}=1.01^{+0.03}_{-0.03}$, and $N_{\rm mos2}=1.03^{+0.03}_{-0.03}$.}
\label{tab:s2_v}
\begin{tabular}{rrrr}\hline\hline
\multicolumn{1}{c}{$kT_1$} & \multicolumn{1}{c}{$kT_2$} & \multicolumn{1}{c}{$kT_3$} & [${\rm keV}$] \\ \hline
$0.21^{+0.05}_{-0.07}$ & \myrule $0.64^{+0.01}_{-0.02}$ & $1.79^{+0.30}_{-0.24}$ & \\ \hline\hline
\multicolumn{1}{c}{$EM_1$} & \multicolumn{1}{c}{$EM_2$} & \multicolumn{1}{c}{$EM_3$} & [$10^{51}\,{\rm cm^{-3}}$] \\ \hline
$2.24^{+1.87}_{-0.60}$ & \myrule $13.84^{+0.82}_{-2.88}$ & $2.88^{+1.05}_{-0.66}$ & \\ \hline\hline
\multicolumn{1}{c}{O} & \multicolumn{1}{c}{Mg} & \multicolumn{1}{c}{Si} & \\ \hline
$0.64^{+0.19}_{-0.14}$ & \myrule $0.27^{+0.12}_{-0.07}$ &$0.47^{+0.16}_{-0.07}$ & \\ \hline\hline
\multicolumn{1}{c}{S} & \multicolumn{1}{c}{Fe} & \multicolumn{1}{c}{Ni} & $\chi^2_{\rm red}$ (dof) \\ \hline
$0.50^{+0.27}_{-0.20}$ & $0.23^{+0.04}_{-0.03}$ & $0.00^{+0.28}_{-0.00}$ & 1.24 (658) \\
\hline\hline
\end{tabular}
\end{center}
\end{table}

We use this spectrum as a baseline for time-resolved spectroscopy. The
EPIC lightcurve of \yygem is split in a total of 15 time intervals
(listed in Table~3) 
\begin{table}
\begin{center}
\caption{Time intervals selected for a systematic investigation of the
evolution of spectral parameters throughout the {\em XMM-Newton}
EPIC observation from 29/30 Sep 2000.} 
\begin{tabular}{lrlrr}\hline
Start & Stop & Remarks & Interval \\ 
\multicolumn{2}{c}{[JD - 2451817]} & & \\ \hline
0.290 & 0.450 & pre-flare quiescence         & $t_1$    \\ 
0.450 & 0.475 & hump before flare            & $t_2$    \\ 
0.475 & 0.484 & rise flare 1                 & $t_3$    \\
0.484 & 0.493 & decay (a) flare 1            & $t_4$    \\ 
0.493 & 0.510 & decay (b) flare 1            & $t_5$    \\ 
0.510 & 0.525 & mini-flare                   & $t_6$    \\
0.525 & 0.630 & post-flare quiescence        & $t_7$    \\
0.630 & 0.662 & secondary eclipse (1st half) & $t_8$    \\ 
0.662 & 0.688 & secondary eclipse (2nd half) & $t_9$    \\
0.688 & 0.710 & post-eclipse feature (a)     & $t_{10}$ \\
0.710 & 0.735 & post-eclipse feature (b)     & $t_{11}$ \\
0.735 & 0.775 & post-eclipse feature (c)     & $t_{12}$ \\
0.775 & 0.790 & rise flare 2                 & $t_{13}$ \\
0.790 & 0.805 & decay flare 2                & $t_{14}$ \\
0.805 & 0.869 & `high state'                 & $t_{15}$ \\
\hline
\end{tabular}
\end{center}
\end{table}
representing different activity levels of the star, and the spectrum
of each phase is modeled by a 3-T model. As the integrated light from the 
quiescent corona should be visible at all times we hold all temperatures
and abundances fixed on the values given in Table~2, and vary only the
emission measure. In some of the time segments, 
namely for the post-eclipse feature and during the large flares,
the 3-T model does not provide an adequate description of the EPIC spectrum:
A high energy excess stands out in the residuals suggesting the presence
of higher temperature material in addition to the emission from the
quiescent corona. Adding a fourth VMEKAL component does not lead to a 
significant improvement. Only a 5-T model represents the data well
($\chi^2_{\rm red} \sim 1$) during the phases of most intense emission. 
For the modeling
of these time intervals we have fixed spectral components \#$\,1-3$ on their
quiescent values (see Table~2). All abundances of components \#\,4 and \#\,5 
have been held fixed on solar values because the statistics do not allow to
constrain further parameters. The last time interval 
($t_{\rm 15}$; the `high state') is
an exception: The signal at high energies is larger than for all other
time segments, and broad Fe K-shell emission is clearly visible (see Fig.~2).
We find an acceptable solution in this case for 
$\frac{\rm Fe}{\rm H} = 0.47^{+0.10}_{-0.09}$.
\begin{figure}
\plottwo{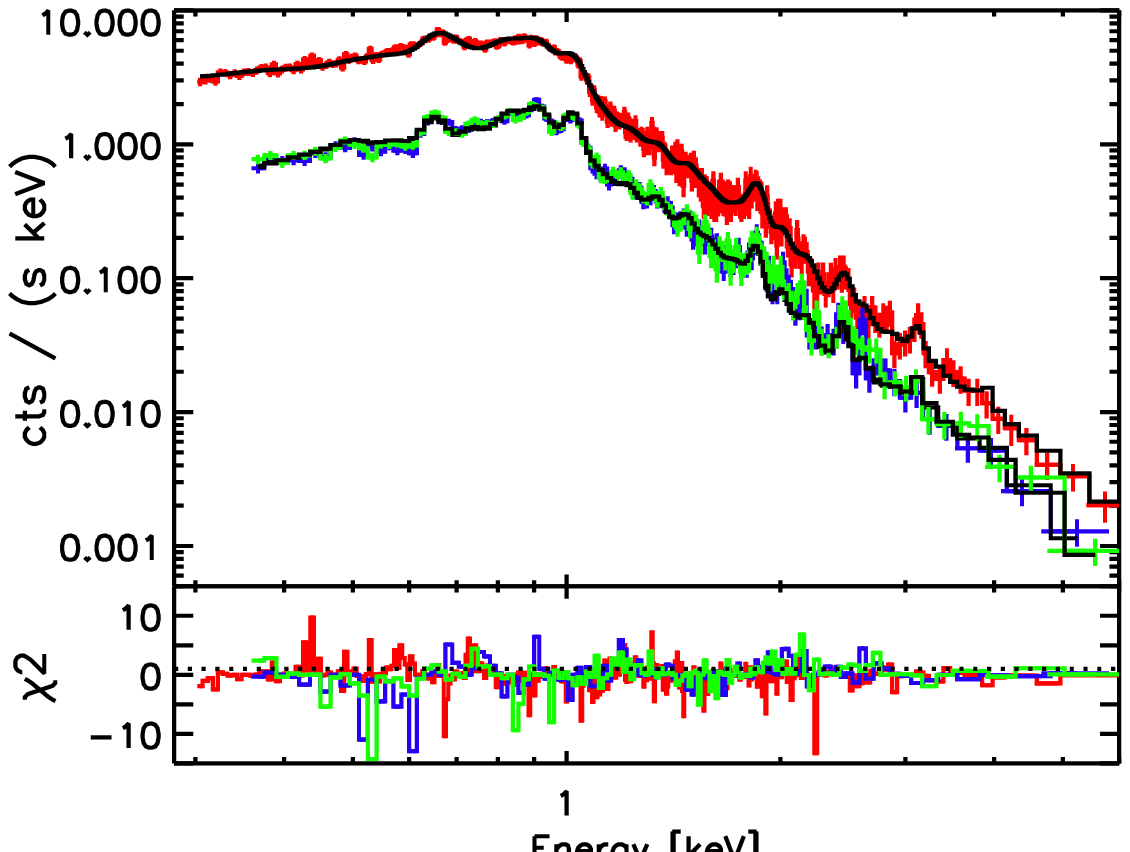}{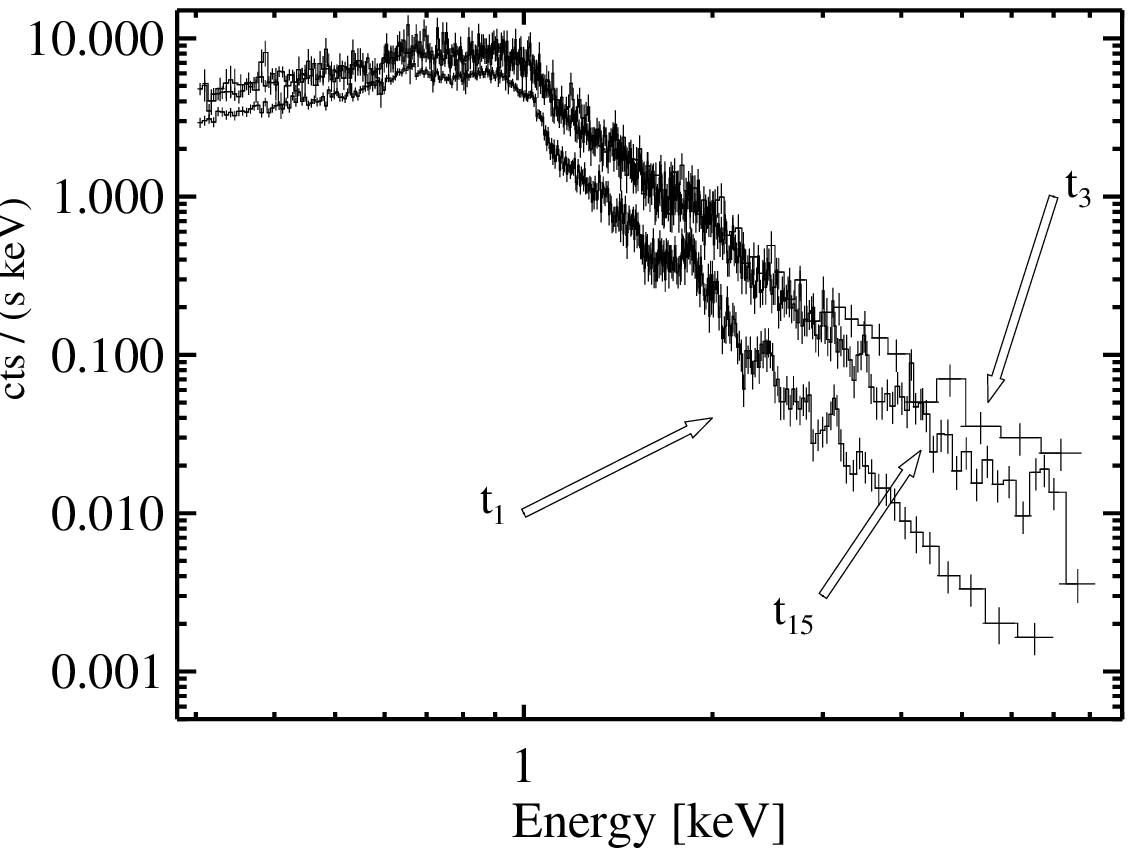}
\caption{{\em XMM-Newton} EPIC spectrum. {\em left} - Quiescent spectrum
($t_1$) of pn, MOS\,1, and MOS\,2 with $\chi^2$ residuals; 
{\em right} - Three different activity
states seen with pn: Quiescence ($t_1$), flare rise ($t_3$), and 
`high state' ($t_{15}$); note the Fe K-shell emission at 
$\sim 6.7$\,keV during $t_{15}$.}
\end{figure}

\subsection{Temperature - Emission Measure Diagrams}

The evolution of temperature and emission measure puts important
constraints on the dynamics during flare decays. In a one-dimensional
hydro-dynamic approach to model stellar flares 
developed by Reale et al. (1993)
the duration of the heating determines the slope in the 
$\lg{T} - \lg{(\sqrt{EM})} - $diagram. 
We have derived $\lg{T} - \lg{(\sqrt{EM})} -$diagrams 
for the spectral components of the 5-T model that represent   
the heated plasma during the two large flares, 
i.e. VMEKAL components \#\,4 and \#\,5.
\begin{figure}
\begin{center}
\plotfiddle{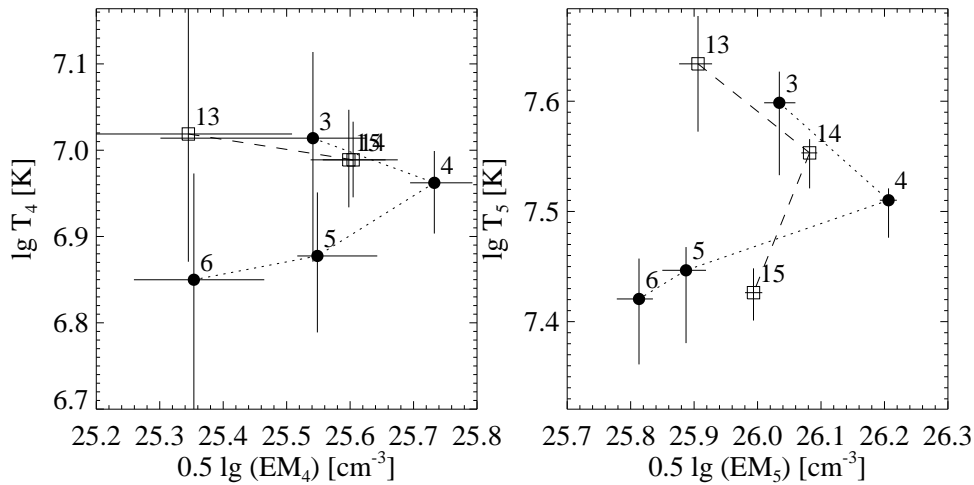}{7cm}{0}{74}{74}{-170}{0}
\caption{Temperature -- emission measure -- diagram for the two flares in the
Sep 2000 observation of YY\,Gem. 
{\em left} - VMEKAL component \# 4, {\em right} -
VMEKAL component \# 5. The numbers next to the data points
indicate the respective time intervals from Table~3.}
\end{center}
\end{figure}
Fig.~3 shows the evolution of the two large flares, both 
starting with the rise phase (time interval $t_3$ and $t_{13}$, respectively).
Under the assumption that the flare emission is concentrated in a single
loop the slope $\zeta$ observed 
during the decay phase can be used to obtain an 
estimate for the loop half-length $L$. This method has been calibrated
for several instruments including EPIC-pn (F. Reale, priv. comm.). 
We apply the equivalent of Eq.~2 from Reale et al. (1997) to derive 
$L$ from the slope $\zeta$, the observed temperature ($T_{\rm max} = 39\,$MK), 
and the decay constant of the lightcurve ($\tau_{\rm lc} = 16 \pm 1$\,min).
The resulting loop length is $L \sim 2 \cdot 10^9$\,cm.

\section{High-resolution Spectra: {\em XMM-Newton} RGS and {\em Chandra} LETGS}

A comparison of the time-averaged 
first order X-ray spectra of \yygem as observed with LETGS and RGS is
given in Fig.~4. We only show the region between $10 - 26$\,\AA, which
contains the strongest lines. Line identifications are given on top
of the diagram. 
\begin{figure}
\begin{center}
\plotfiddle{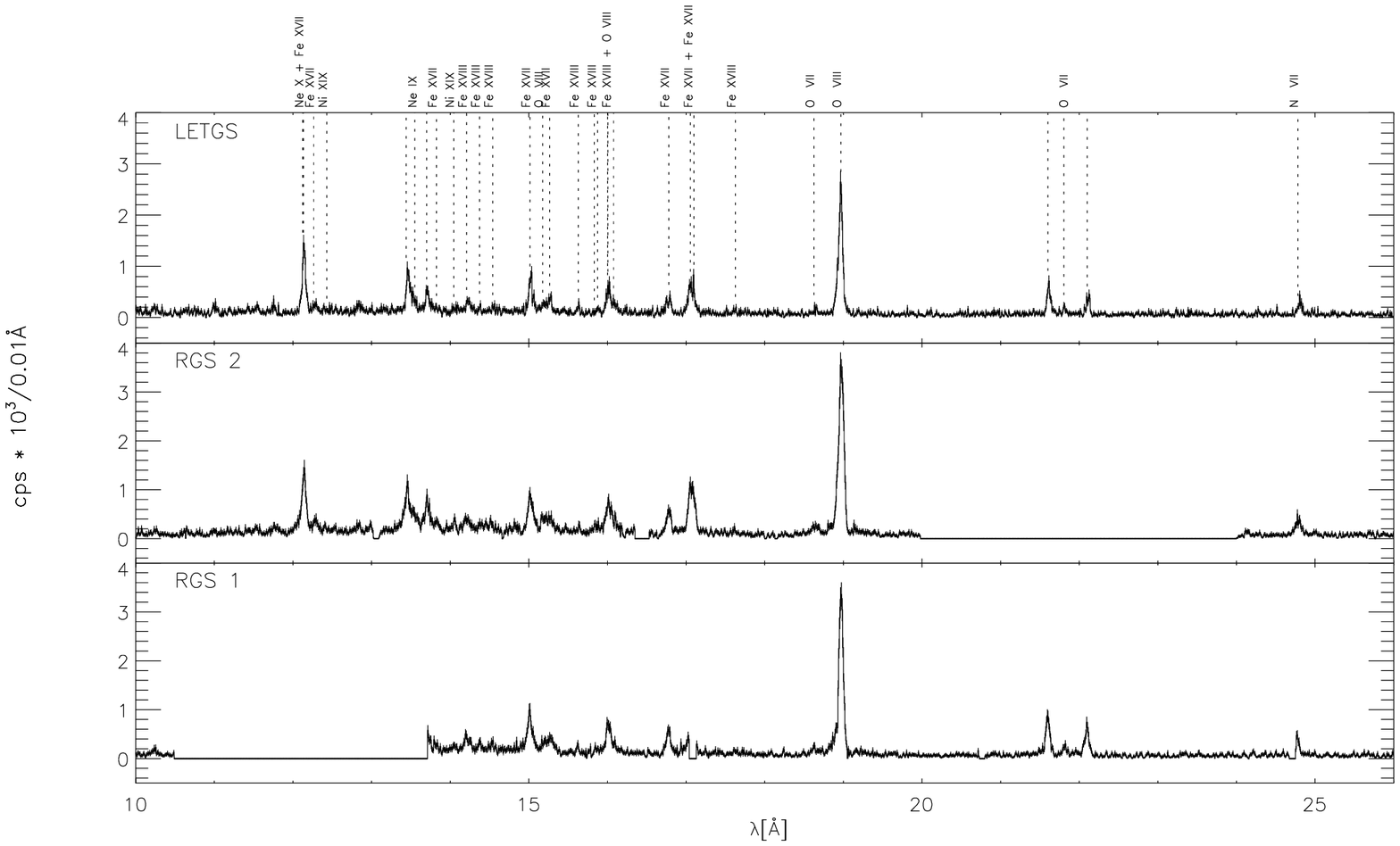}{10cm}{270}{90}{90}{-220}{340}
\end{center}
\vspace*{8.5cm}
\caption{Comparison of the simultaneous {\em XMM-Newton} RGS\,1 and
RGS\,2, and {\em Chandra} LETGS spectrum of \yygem in the range $\lambda =
10-26$\,\AA. Straight horizontal lines in the RGS spectrum represent data gaps
due to CCD failure or gaps inbetween individual chips.}
\end{figure}
The spectrum is given in units of cts/s/bin. Since the RGS and the LETGS
observations overlap for about $75$\,\% in time, the relative strength of
the lines measured by both instruments should be similar, with some 
dependence of the line strength on the binsize, and the
absolute numbers demonstrate directly the difference in sensitivity 
between RGS and LETGS.

The Ly$\alpha$ line of H-like O\,VIII is by far the strongest line 
in the spectrum with the highest photon flux, 
i.e. taking account of the effective area. 
Next to a number of iron L-shell transitions we
identify the He-like triplets of four elements: Si\,XIII, 
Ne\,IX, O\,VII, and N\,VI.
The O\,VII triplet is the strongest triplet and the
only one which is clearly resolved and not blended with other lines.
A detailed investigation of the properties of the coronal plasma 
making use of line ratios will be presented by Stelzer et al., in prep.

\section{Summary}

The X-ray lightcurve of \yygem shows that the object was subject to strong
variability including two large outbursts during the time of observation.
The parameters of a 3-T model for the quiescent emission
are compatible with results from the analysis of an earlier {\em
XMM-Newton} observation of YY\,Gem presented by G\"udel et al. (2001).
Time-resolved modeling of the EPIC spectrum reveals the presence of
a high temperature plasma ($kT_{\rm max} = 3.4\,$keV) in flares.  
According to a one-dimensional hydrodynamic model the flare emission
arises in a semi-circular loop with $\sim 2 \cdot 10^9$\,cm length. 
This approach is certainly a simplification of the real situation which does 
involve a multi-temperature plasma and possibly complex loop systems. 
Nevertheless, the hydrodynamic approach is important: while simple 
quasi-static modeling tends to reproduce large loops the method applied here 
demonstrates that the coronal structures are likely to be much smaller 
than the radii of both stars in the \yygem system. 

The simultaneous observation of \yygem with {\em Chandra} and 
{\em XMM-Newton} demonstrates the different sensitivity of these 
instruments. Each of the two RGS provides roughly the same count rate
as the LETGS first order spectrum. The LETGS is more sensitive at
short wavelengths (see e.g. the region around the Ne\,IX triplet),
while the sensitivity of RGS is slightly higher towards longer
wavelengths (e.g. near the O\,VIII Ly$\alpha$ line).

\acknowledgements{This work is supported by the Bundesministerium f\"ur
Bildung und Forschung/ Deutsches Zentrum f\"ur Luft- und Raumfahrt 
(BMFT/ DLR), the Max-Planck Society, and the Heidenhain-Stiftung. MA
acknowledges support from the Swiss National Science Foundation 
(grants 2100-049343).}

\end{document}